\documentclass{PoS}

\title{Exotic hadron holography from anomalous dimensions}

\ShortTitle{Exotic hadron holography from anomalous dimensions}

\author{\speaker{Hilmar Forkel}\\
        Institut f\"ur Physik, Humboldt-Universit\"at zu Berlin, D-12489 Berlin, Germany \\
        E-mail: \email{forkel@physik.hu-berlin.de}}

\abstract{The anomalous dimensions of hadronic interpolators contain dynamical information on the properties of the associated hadron states. We point out that they provide, in particular, a link by which  gauge-invariant information on exotic contributions to hadronic wavefunctionals can be obtained from approximate gravity duals for QCD. This is demonstrated by the holographic description of a dominant tetraquark component in the lightest scalar mesons. 
}

\FullConference{Xth Quark Confinement and the Hadron Spectrum,\\
		October 8-12, 2012\\
		TUM Campus Garching, Munich, Germany}

\begin{document}

While the anomalous dimensions of color-singlet operators play a central role in the original, conformal versions of the gauge/string correspondence \cite{revs1}, they have only recently begun to enter holographic approaches to QCD. In particular, anomalous dimensions of hadronic interpolators were implemented to provide an AdS/QCD   \cite{revs2}  description of quark correlations inside hadrons \cite{for09,for10} which can have a significant and in exotic cases even striking impact on the hadron properties. Although multiquark components in hadronic wave functionals are typically gauge dependent, a holographic description is still possible because  the five-dimensional bulk modes are dual to the interpolators of the corresponding hadrons. Hence the anomalous dimensions of these interpolators import gauge-invariant information on their quark content and couplings, and thus on the multiquark correlations in the corresponding hadron, as bulk-mode mass corrections into the gravity dual. 

This AdS/QCD representation of multiquark effects was originally introduced to describe diquark correlations in baryons  \cite{for09}. It changes the resulting light-quark baryon excitation spectrum into
\begin{equation}
M_{n,L}^{2}=4\lambda ^{2}\left(n+L+\frac{3}{2}\right) -2\left( M_{\Delta
}^{2}-M_{N}^{2}\right) \kappa   \label{dcbms}
\end{equation}
(where $\lambda$ is the IR scale of the \textquotedblleft metric
soft-wall\textquotedblright\ gravity dual \cite{for07} 
while $n$ ($L$) denotes the radial (angular momentum) excitation 
level). The second term, proportional to the baryon's \textquotedblleft good-diquark fraction\textquotedblright\ $\kappa$, is generated by suitable anomalous dimensions for the QCD nucleon interpolators. Equation (\ref{dcbms})  describes the linear square-mass trajectories of the over 40 measured nucleon and delta (with $\kappa=0$) resonances with unprecendented accuracy. The dual mode solutions further reveal  that baryons with larger $\kappa$ have a smaller size.

Encouraged by these results, the anomalous-dimension-induced representation of multiquark correlations was then applied to the more challenging holographic description of exotic hadrons with a non-standard (valence) quark content. The light scalar meson sector  \cite{clo02} with its expected  tetraquark component \cite{jaf77} was examined in Ref. \cite{for10}. The radial bulk equation for the modes dual to the scalars can be written as the Sturm-Liouville problem 
$
\left[ -\partial _{z}^{2}+V\left( z\right) \right] \phi \left( q,z\right)
=q^{2}\phi \left( q,z\right)  \label{sleq}
$. In the dilaton soft-wall gravity dual \cite{kar06} without anomalous-dimension contributions, the potential $V$ has the form
\begin{equation}
V\left( z\right) =\left( \frac{15}{4}+m_{5}^{2}R^{2}\right) \frac{1}{z^{2}}
+\lambda ^{2}\left( \lambda ^{2}z^{2}+2\right).   \label{vsw}
\end{equation}
The anomalous 
dimension $\gamma(z)$ of the tetraquark interpolator $J_{\bar{q}^{2}q^{2}}$  (i.e. the local four-quark operator which most strongly couples to the tetraquark state)  with scaling dimension $\Delta _{\bar{q}^{2}q^{2}}=6+\gamma(z)$ adds the universal contribution 
\begin{equation}
\Delta V\left( z\right) =\gamma \left( z\right) \left[ \gamma \left(
z\right) +8\right] \frac{1}{z^{2}}  \label{DelV}
\end{equation}
to the potential (\ref{vsw}) with $m_{5}^{2}R^{2}=12$.  Eq. (\ref{DelV}) implies the  crucial lower bound $\Delta V\left(z\right) \geq -16/z^{2}$ which holds for any $\gamma $ and prevents the collapse of the dual modes into the AdS$_{5}$ boundary. This bound is saturated by $\gamma \equiv -4$ and therefore determines the lightest tetraquark mass
\begin{equation}
M_{\bar{q}^{2}q^{2},0}\geq M_{\Delta =2,0}=2\lambda  \label{mbd}
\end{equation}
which the anomalous-dimension-induced holographic binding mechanism can produce. Moreover, for constant values $-4< \gamma <-3$ the tetraquark ground state is lighter than its $\bar{q} q$ counterpart. Since $\gamma $ only enters through the mass term of the bulk mode which is model-independently prescribed by the AdS/CFT dictionary, the correction (\ref{DelV}) and the associated binding mechanism will arise in other AdS/QCD duals as well.

To estimate the quantitative impact of the anomalous-dimension contribution $\Delta V$ (until direct QCD information on the RG flow of $\gamma $ will eventually become available and fix $\Delta V$ uniquely), a typical power ansatz $\gamma \left( z\right) =-az^{\eta }+bz^{\kappa }$ can be adopted. Its coefficients turn out to be tightly constrained by consistency and stability requirements but can still  produce almost maximal ground-state binding \cite{for10}. The latter drives the mass $M_{\bar{q}^{2}q^{2},0}$ of the lightest tetraquark from $\sim 40\%$ above (for $\gamma  \equiv 0$) down to $\sim 20\%$ below the $\bar{q}q$ ground-state mass $M_{q\bar{q},0}
=\sqrt{6}\lambda $. The resulting  masses $M_{\bar{q}^{2}q^{2},n}$ of the tetraquark excitations get pushed beyond the corresponding $M_{\bar{q}q,n}$ from around $n\gtrsim 2$.  The higher-lying radial 
tetraquark excitations will therefore likely be broad enough to prevent the appearance of supernumeral states in the scalar meson spectrum. 

It should be interesting to extend the anomalous-dimension-based holographic description of non-valence quark components to other exotics, including heavy tetraquarks, pentaquarks and hybrids. Moreover, anomalous-dimension-induced corrections also encode other aspects of hadronic structure which largely remain to be explored.

\acknowledgments{It is a pleasure to thank the organizers for a very informative and enjoyable conference. }


\begin{thebibliography}{99}

\bibitem{revs1} O. Aharony et al., \emph{Large-N field theories, string theory and gravity}, Phys. Rep. \textbf{323}  (2000) 183.

\bibitem{revs2}
Y. Kim and D. Yi, \emph{Holography at work for nuclear and hadron physics}, Adv. High Energy Phys. \textbf{2011} (2011) 259025;  S.J. Brodsky and G.F. de T\'{e}ramond, \emph{AdS/CFT and Light-Front QCD}, \texttt{arXiv:0802.0514}.

\bibitem{for09} H. Forkel and E. Klempt, \emph{Diquark correlations in
baryon spectroscopy and holographic QCD}, Phys. Lett. \textbf{B 679}
(2009) 77.

\bibitem{for10} H. Forkel, \emph{Light scalar tetraquarks from a holographic 
perspective}, Phys. Lett. \textbf{B 694} (2010) 252; 
\emph{Multiquark correlations in light mesons and baryons from holographic QCD},
AIP Conf. Proc. \textbf{1388} (2011) 182 [\texttt{arXiv:1103.3902}].

\bibitem{for07} H. Forkel, M. Beyer and T. Frederico, \emph{Linear
square-mass trajectories of radially and orbitally excited hadrons in
holographic QCD}, \emph{JHEP} \textbf{07} (2007) 077; \emph{Linear 
meson and baryon trajectories in AdS/QCD}, \emph{Intl. J.Mod. Phys. E} 
\textbf{16}\ (2007) 2794.

\bibitem{clo02} 
E. Klempt and A. Zaitsev, \emph{Glueballs, hybrids, multiquarks -- Experimental facts versus QCD inspired concepts}, Phys. Rep. \textbf{454}, 1 (2007); C. Amsler and N.A. T\"{o}rnqvist, \emph{Mesons beyond the naive quark model}, Phys. Rep. 
\textbf{389}, 61 (2004); D.V. Bugg, \emph{Four sorts of meson}, Phys. Rept. \textbf{397}, 257 (2004);  E. Ruiz Arriola and W. Broniowski, \emph{Scalar-isoscalar states in the large-$N_c$ Regge approach}, 
Phys. Rev. D. \textbf{81}  (2010) 054009.

\bibitem{jaf77} R.L. Jaffe, \emph{Phenomenology of 
$ Q^2\bar{Q}^2$ mesons}, Phys. Rev. D \textbf{15}, 267, 281 (1977); 
J.R. Pel\'{a}ez, \emph{Nature of light scalar mesons from their large-N$_c$ behavior}, Phys. Rev. Lett. \textbf{92} (2004) 102001.

\bibitem{kar06} A. Karch, E. Katz, D.T. Son and M.A. Stephanov, \emph{Linear Confinement and AdS/QCD}, Phys. Rev. D \textbf{74}  (2006) 015005.


\end{thebibliography}
\end{document}